\documentclass[prl,superscriptaddress,twocolumn,longbibliography]{revtex4-1}

\usepackage{amsmath}
\usepackage{amssymb}
\usepackage{amsxtra,color}
\usepackage{latexsym}
\usepackage{graphicx}
\usepackage{tikz}
\usepackage{pgfplots}
\usepgfplotslibrary{groupplots}
\pgfplotsset{compat=1.3}
\usepackage{subcaption}
\usepackage{lipsum}
\usepackage{color}
\usepackage{braket}
\usepackage{lineno}

\usepackage[utf8]{inputenc}
\usepackage{MnSymbol}	

\usepackage{hyperref}

\definecolor{MyDarkGreen}{rgb}{0,0.6,0}
\definecolor{MyDarkBlue}{rgb}{0,0,0.8}
\definecolor{MyDarkRed}{rgb}{0.6,0,0.3}

\hypersetup{breaklinks=true, colorlinks=true,plainpages=true, linktocpage=true, linkcolor=MyDarkBlue, citecolor=MyDarkGreen, urlcolor=MyDarkRed, pdfborder={0 0 0},%
pdfauthor={},%
pdfsubject={Research article},
pdftitle={}%
}

\begin{document}

\title{Probing electronic decoherence with high-resolution attosecond photoelectron interferometry}

\author{D. \surname{Busto}}
\email{david.busto@fysik.lth.se}
\affiliation{Department of Physics, Lund University, Box 118, 22100 Lund, Sweden}
\affiliation{Institute of Physics, Albert Ludwig University, Stefan-Meier-Strasse 19, 79104  Freiburg, Germany}

\author{H. \surname{Laurell}}
\affiliation{Department of Physics, Lund University, Box 118, 22100 Lund, Sweden}

\author{D. \surname{Finkelstein Shapiro}}
\affiliation{Department of Chemical Physics, Lund University, Box 124, 22100 Lund, Sweden}

\author{C. \surname{Alexandridi}}
\affiliation{Universit\'e Paris-Saclay, CEA, CNRS, LIDYL, 91191 Gif-sur-Yvette, France}


\author{M. \surname{Isinger}}
\affiliation{Department of Physics, Lund University, Box 118, 22100 Lund, Sweden}

\author{S. \surname{Nandi}}
\affiliation{Department of Physics, Lund University, Box 118, 22100 Lund, Sweden}

\author{R. J. \surname{Squibb}}
\affiliation{Department of Physics, University of Gothenburg, Origov\"agen 6B, 41296 Gothenburg, Sweden}

\author{M. \surname{Turconi}}
\affiliation{Universit\'e Paris-Saclay, CEA, CNRS, LIDYL, 91191 Gif-sur-Yvette, France}

\author{S. \surname{Zhong}}
\affiliation{Department of Physics, Lund University, Box 118, 22100 Lund, Sweden}

\author{C. L. \surname{Arnold}}
\affiliation{Department of Physics, Lund University, Box 118, 22100 Lund, Sweden}

\author{R. \surname{Feifel}}
\affiliation{Department of Physics, University of Gothenburg, Origov\"agen 6B, 41296 Gothenburg, Sweden}

\author{M. \surname{Gisselbrecht}}
\affiliation{Department of Physics, Lund University, Box 118, 22100 Lund, Sweden}

\author{P. \surname{Sali\`{e}res}}
\affiliation{Universit\'e Paris-Saclay, CEA, CNRS, LIDYL, 91191 Gif-sur-Yvette, France}

\author{T. \surname{Pullerits}}
\affiliation{Department of Chemical Physics, Lund University, Box 124, 22100 Lund, Sweden}

\author{F. \surname{Mart\'in}}
\affiliation{Departamento de Qu\'imica, Modulo 13, Facultad de Ciencias, Universidad Aut\'onoma de Madrid, 28049 Madrid, Spain}
\affiliation{Instituto Madrileno de Estudios Avanzados en Nanociencia (IMDEA-Nanoscience), Cantoblanco, 28049 Madrid, Spain
}
\affiliation{Condensed Matter Physics Center (IFIMAC), Universidad Aut\'onoma de Madrid, 28049 Madrid, Spain}

\author{L. \surname{Argenti}}
\affiliation{Department of Physics, University of Central Florida, Orlando, Florida 32816, USA}
\affiliation{CREOL, University of Central Florida, Orlando, Florida 32816, USA}

\author{A. \surname{L'Huillier}}
\affiliation{Department of Physics, Lund University, Box 118, 22100 Lund, Sweden}

\pacs{}

\begin{abstract}
Quantum coherence plays a fundamental role in the study and control of  ultrafast dynamics in matter. In the case of photoionization, entanglement of the photoelectron with the ion is a well known source of decoherence when only one of the particles is measured. Here we investigate decoherence due to entanglement of the radial and angular degrees of freedom of the photoelectron. We study two-photon ionization via the 2s2p autoionizing state in He using high spectral resolution photoelectron interferometry. Combining experiment and theory, we show that the strong dipole coupling of the 2s2p and 2p$^2$ states results in the entanglement of the angular and radial degrees of freedom. This translates, in angle integrated measurements, into a dynamic loss of coherence during autoionization.
\end{abstract}

\maketitle 
The development of attosecond light sources since the beginning of the 21$^\text{st}$ century has opened the possibility to probe electronic dynamics with attosecond resolution. The absorption of an attosecond light pulse by matter leads to the emission of a photoelectron wavepacket (EWP) corresponding to the coherent superposition of continuum states. The measurement of the amplitude and phase of the EWPs, using attosecond streaking \cite{KienbergerNature2004} or the reconstruction of attosecond beating by interference of two-photon transitions (RABBIT) technique \cite{PaulScience2001} provides information on the photoionization dynamics in atoms and molecules in the gas phase \cite{SchulzeScience2010,IsingerScience2017,Zhong2020,HaesslerPRA2009,Nandi2020,HuppertPRL2016,CattaneoNP2018,KamalovPRA2020} as well as in solids \cite{CavalieriNature2007,KasmiOptica2017} and liquids \cite{JordanScience2020}. 
One of the successful applications of the RABBIT technique has been the study of the ionization dynamics close to an autoionization resonance \cite{GrusonScience2016,Busto2018,Barreau2019,KoturNC2016,Cirelli2018,Turconi2020}.
The interpretation of these measurements relies on the assumption that the EWPs are fully coherent and can be represented by a complex-valued wavefunction, \textit{i.e.} a pure quantum state. 

In general, however, quantum experiments only probe a reduced number of degrees of freedom. As a result, measurements are averaged over the degrees of freedom outside of the studied subsystem. 
If the subsystem is entangled with some of the degrees of freedom that are not measured, averaging over the latter leads to a loss of coherence in the measurements. A typical example is the sudden excitation of  a coherent superposition of electronic states in a molecule. The coupling of the molecular electron wavepacket with the vibrational degrees of freedom results in a loss of its coherence on a time scale ranging from a few to hundreds of femtoseconds  when the vibrational degrees of freedom are not accessible in the measurement \cite{Arnold2017,Vacher2017,LaraAstiaso2016}. 

Quantifying decoherence is a difficult task,  often overlooked in attosecond photoionization experiments since coupling to the outside environment usually happens on a time scale much longer than that of the photoionization process. Recently, a few studies have investigated the loss of coherence due to incomplete measurements, \textit{i.e.} not probing all degrees of freedom of the photoelectron and the ion. For example, the entanglement between the photoelectron and the ion leads to a loss of coherence when only one of the two particles is measured \cite{Pabst2011,Arnold2020,Vrakking2021,KollArxiv2021,NishiPRA2019}. The loss of purity in photoelectron interferometry, due to experimental fluctuations and limited spectral resolution, has been investigated by quantum state tomography methods  \cite{BourassinBouchet2015,BourassinBouchet2020}.


As demonstrated by B\"arnthaler et al. for microwave cavities ~\cite{Baernthaler2010}, Fano resonances are sensitive probes of decoherence: since they result from an interference phenomenon, any small perturbation is immediately apparent as a loss of contrast and reduced phase variation across the resonance.  In the case of photoionization, the interference takes place between direct ionization of an atom or molecule in its ground state and autoionization from a quasi-bound state embedded in the continuum \cite{FanoPR1961}. Perturbations to this scheme, such as the influence of additional levels, the interaction with a thermal bath, and/or experimental imperfections such as laser jitter \cite{FinkelsteinPRA2016,FinkelsteinPRA2018} may lead to decoherence. A high spectral resolution is generally needed to identify experimentally the modifications of the resonance profiles due to decoherence.


In this work, using an infrared (IR) field with a narrow spectral bandwidth, together with a deconvolution algorithm to compensate for the electron spectrometer's response function, we perform interferometric photoelectron spectroscopy measurements, using the RABBIT technique, with unprecedented spectral resolution. We determine the amplitude and phase of EWPs created by resonant absorption of extreme ultraviolet (XUV) radiation close to the 2s2p resonance in He, plus absorption or stimulated emission of an IR photon, as shown in Fig.~\ref{Fig1}(a). The high spectral resolution of our measurements allows us to investigate the degree of coherence of the emitted EWPs and to show that it depends on whether the IR photon is absorbed or emitted during the two-photon transition. The coupling of 2s2p to the 2p$^2$ ($^1$S) autoionizing state, which is energetically close in the absorption case, induces a loss of purity in the measured angle-integrated wavepacket. This is verified by comparing with the two-photon wavepacket originating from emission of an IR photon from the 2s2p resonance, which does not suffer from the presence of resonances in the final state and acts as a benchmark.

\section{Results}

\subsection{Experimental results}
\begin{figure}
    \includegraphics[width=\columnwidth]{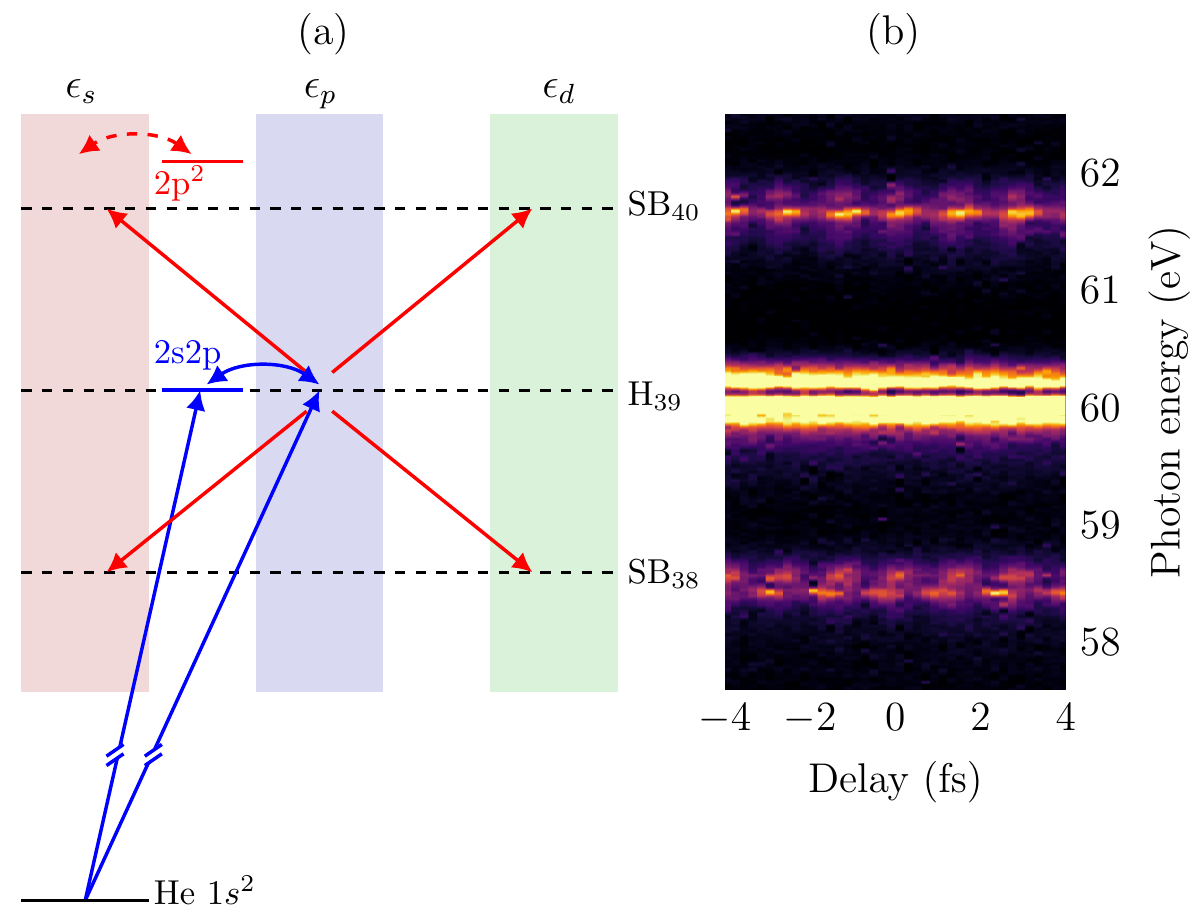}
    \caption{Principle of the experiment. (a) Spectroscopic scheme of the states and transitions relevant to this study. The blue/red arrows indicate XUV/IR dipole transitions. The double headed arrows indicate the configuration interaction between the bound states and the continuum. The reference non-resonant transitions involving H$_{37}$ and H$_{41}$, are not shown for simplicity. (b) RABBIT spectrogram after deconvolution.}
    \label{Fig1}
\end{figure}

In our experiment, high-order harmonics are generated by focusing a 30~fs IR pulse in a cell filled with neon gas, resulting in the emission of an XUV comb of odd harmonics of the laser central frequency. The central wavelength of the IR field is chosen so that the 39\textsuperscript{th} harmonic is resonant with the 2s2p resonance in He, which is located  at $60.147$~eV above the ground state \cite{Madden1963} [see Fig.~\ref{Fig1}(a)]. A 2m-long magnetic bottle electron spectrometer (MBES), with a spectral resolution below 100~meV in the 0-5 eV spectral range, is used to detect the photoelectrons. To benefit from this resolution, a retarding voltage is applied so that electrons created by absorption of the 39\textsuperscript{th} harmonic and the adjacent sidebands called SB$_{38}$ and SB$_{40}$ are in the 0-5~eV range. SB$_{2q}$ originates from the interference of two quantum paths: absorption of harmonic $2q+1$ and emission of an IR photon or absorption of harmonic $2q-1$ and an IR photon. The spectral resolution of our measurements is further improved using a blind Lucy-Richardson deconvolution algorithm \cite{Richardson1972,Lucy1974}. In a traditional RABBIT setup, both the XUV and IR pulses have a broad bandwidth. Consequently, several combinations of XUV and IR frequencies lead to the same final energy and interfere \cite{JimenezGalan2016}. This finite pulse effect induces distortions of resonant two-photon ionization spectra and thus a loss of spectral resolution. To minimize this effect, the spectral bandwidth of the probe pulse is reduced to 10~nm (full width at half maximum) using a band-pass filter. By comparison, the bandwidth of the IR pulse used to generate the harmonics is approximately 50~nm. 

The measured RABBIT spectrogram is shown in Fig.~\ref{Fig1}(b). The photoelectron spectrum corresponding to the absorption of harmonic 39 and the two neighbouring sidebands present a double structure, with an interference minimum, owing to the resonance. When the delay $\tau$ between the XUV and IR pulses is varied, the sideband signal oscillates as:
\begin{equation}
\begin{split}
I^{(\pm)}(E,\tau)=&|A_\mathrm{R}(E)|^2+|A_\mathrm{NR}(E)|^2\\
+&2|A_\mathrm{R}(E)||A_\mathrm{NR}(E)|\cos[2\omega\tau\mp\Delta\phi(E)],
\label{Eq:RABBIT}
\end{split}
\end{equation}
where $\omega$ is the central angular frequency of the IR pulse, $E$ is the net photon energy absorbed, $A_\mathrm{R}(E)$ and $A_\mathrm{NR}(E)$ are respectively the resonant and non-resonant two-photon transition amplitudes, and $\Delta\phi(E)=\arg(A_\mathrm{R})-\arg(A_\mathrm{NR})$ is the phase difference between the two quantum paths. The sign $\pm$ indicates if the sideband is above $(+)$ or below $(-)$ the resonant harmonic (H$_{39}$). In the Rainbow version of the RABBIT method \cite{GrusonScience2016}, the oscillations of the sidebands are fitted with Eq.~(\ref{Eq:RABBIT}) for each final energy $E$, allowing the extraction of the amplitude and phase as a function of energy, thus mapping the resonance structure. 


\begin{figure}
    \centering
    \includegraphics[width=\columnwidth]{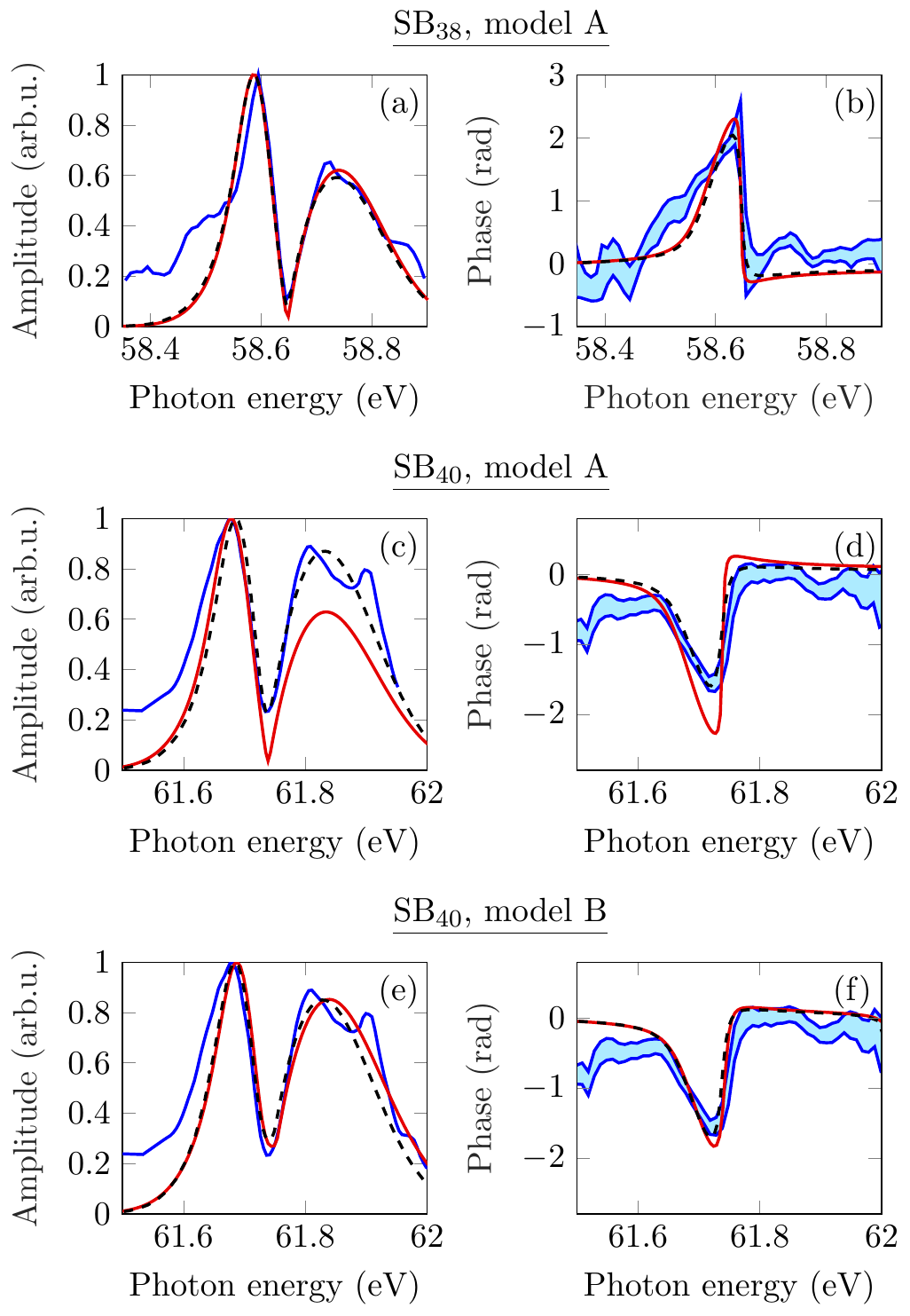}
        \caption{Amplitude (a,c,e) and phase difference, $\Delta\phi(E)$, (b,d,f) measured in sidebands SB$_{38}$~(a,b) and SB$_{40}$~(c-f). The experimental measurements are shown in blue, the results of the fits are shown in dashed black and the results of the fully coherent models in red [model A in (a-d) and model B in (e,f), see main text]. The blue shaded areas in (b,d,f) correspond to the standard deviation around the measured value.}
    \label{Fig2}
\end{figure}

Figure \ref{Fig2} shows in blue the amplitude and phase measured in SB$_{38}$ (a,b) and in SB$_{40}$ (c,d). In SB$_{38}$, the amplitude exhibits a sharp interference structure which goes almost to zero at $\sim 58.65$~eV (a) with a sharp phase jump of approximately 2 rad (b). In SB$_{40}$, the interference contrast is reduced (c) and the phase variation is smoother, over approximately 1.5 rad (d). Compared to previous measurements reported in the literature \cite{GrusonScience2016,Busto2018}, the high spectral resolution of our measurements allows us to measure a significantly higher interference contrast and a larger phase jump. In addition, a clear difference between the two adjacent sidebands is observed, in contrast to previous results \cite{GrusonScience2016,Busto2018}. 

\begin{figure}
    \centering
    \includegraphics[scale=0.9]{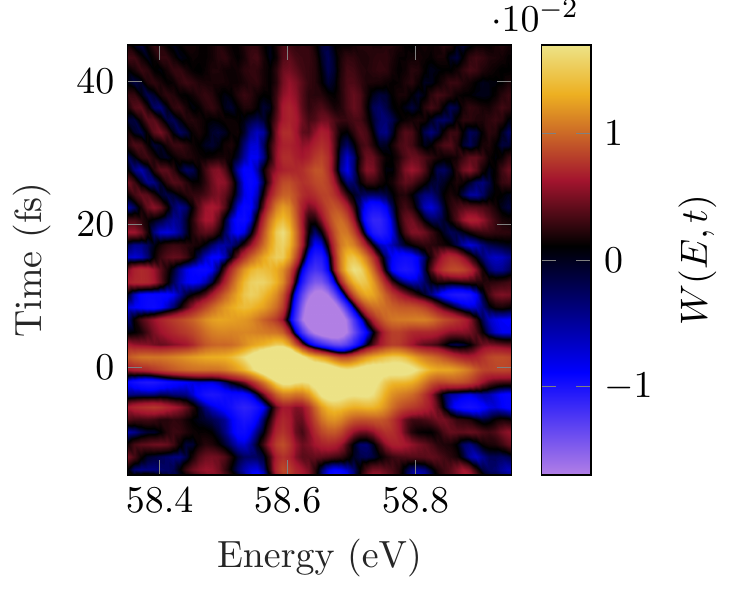}
        \caption{Wigner distribution calculated using the experimental complex amplitude measured in SB$_{38}$. The observation of hyperbolic interference fringes demonstrates the high spectral resolution of our measurements.}
    \label{Fig3}
\end{figure}

Fig.~\ref{Fig3} shows the Wigner distribution for SB$_{38}$, defined as 
\begin{equation}
    W(E,t)=\int A_\mathrm{R}(E-\varepsilon/2)A_\mathrm{R}^*(E+\varepsilon/2)e^{\text{i}\varepsilon t/\hbar} \mathrm{d}\varepsilon,
\end{equation}
where $\hbar$ is the reduced Planck constant. A spectrally broad and temporally narrow feature can be observed around $t=0$, which corresponds to the direct ionization from the ground state to the continuum. The spectrally narrow and temporally long feature around 58.6~eV arises from the decay of the 2s2p resonance in the continuum. Finally, the large minimum at 58.65~eV and the hyperbolic fringes originate from interference between the two ionization paths. The observation of the hyperbolic interference fringes is possible thanks to the high spectral resolution achieved in this work. We note that our definition of the Wigner distribution assumes that the wavepacket, described by the complex amplitude $A_R(E)$, is in a pure state. In the following, we investigate in detail the validity of this assumption.

\subsection{Theoretical calculations}

The measurements are compared to theoretical calculations based on the finite-pulse two-photon resonant model from Refs.~\cite{JimenezPRL2014,JimenezGalan2016}, in which dipole couplings between unperturbed continua are calculated using the on-shell approximation. Autoionization of the 2s2p resonance reached by absorption of a single XUV photon is formally described according to Fano's formalism \cite{FanoPR1961}, with the help of a complex resonance factor defined as
\begin{equation}
    R(\epsilon)=\frac{q+\epsilon}{\epsilon+\text{i}}.
    \label{Eq:FanoFactor}
\end{equation}
The quantity $\epsilon=2(E-E_r)/\Gamma$ is the reduced energy ($E_r$ is the resonance energy and $\Gamma$ the spectral width of the resonance), while $q$ is a (real) asymmetry parameter depending on the relative strength between direct ionization and autoionization (see Fig.~\ref{Fig1}). Bärnthaller \textit{et al.} \cite{Baernthaler2010}, generalizing previous work on quantum dots \cite{ClerkPRL2001}, show that the influence of decoherence on a Fano lineshape can be described using Eq.~(\ref{Eq:FanoFactor}) by allowing $q$ to be complex. The origin of the decoherence, for example dissipation or dephasing, appears in the type of trajectory traced by the $q$ parameter in the complex plane as the degree of coherence is varied.


As described in detail in the Supplementary Material (SM), we use the analytical formulation of \cite{JimenezGalan2016} to describe the two-photon transitions. Although the transition amplitudes cannot be parametrized by a simple resonance factor as in Eq.~(\ref{Eq:FanoFactor}), they can be expressed as a function of the one-photon $q$ parameter, which is real in the absence of coupling to degrees of freedom outside the system considered in the model. 
We use two analytical models  (denoted A and B), including the 2s2p state only (A) or both the 2s2p and the 2p$^2$ state (B). Our approach consists in fitting the experimental data with the two models using as fit parameter a complex-valued $q$ and compare the fitted amplitude and phase to the predictions of the model for a real $q$.

In model A, we assume that the two-photon transition leads to a flat continuum with only a single open channel.  We also neglect polarization effects due to virtual excitations of the intermediate autoionizing state. The results of the fit are shown as dashed black curves in Fig.~\ref{Fig2}(a-d) and the values of $q$ retrieved from the fit are shown in Table \ref{tab:my_label}A. In the case of SB$_{38}$, the $q$ parameter giving the best fit to the experimental data has a small imaginary part and a real part that is in good agreement with single-photon measurements using synchrotron radiation ($q=-2.77$) \cite{DomkePRA1996}. In contrast, for SB$_{40}$, the fitted $q$ parameter has an imaginary part twice as large as for SB$_{38}$ and a real part that significantly deviates from spectroscopic data. We also show in Fig.~\ref{Fig2}(a,b,c,d) the results from model A, assuming no decoherence,  using a real asymmetry parameter $q=-2.77$ (red solid line). This calculation reproduces well the experimental measurements for SB$_{38}$ but fails to describe the reduced contrast and phase variation in SB$_{40}$. The good agreement obtained for SB$_{38}$ allows us to conclude that, in our experimental conditions, decoherence due to limited spectral resolution or experimental fluctuations is negligible. As a consequence, the large deviation of the fitted $q$ parameter in SB$_{40}$, which is measured in the same conditions, cannot be attributed to experimentally induced decoherence. In both sidebands, the deviation between the experimental and theoretical amplitude at low energy can be attributed to a well known distortion introduced by the MBES \cite{KruitJPE1983,Mucke2012,Namba2015}.

\begin{table}[h]
    \centering
    \begin{tabular}{|c|c|c|c|c|}
    \hline
        \ & \multicolumn{2}{|c|}{ Model A: 2s2p} &\multicolumn{2}{|c|}{ Model B: 2s2p and 2p$^2$}\\
        \hline
        \ & $Re(q)$ & $Im(q)$ & $Re(q)$ & $Im(q)$\\
        \hline
        SB$_{38}$ & $-2.6\pm0.05$& $0.14\pm0.06$ & $-2.6\pm0.05$&  $0.14\pm0.06$\\
        \hline
        SB$_{40}$ & $-1.88\pm0.03$&$0.32\pm0.06$ & $-2.72\pm0.02$&$0.16\pm0.04$\\
        \hline
    \end{tabular}
    \caption{Values of the real and imaginary values of the $q$ parameter retrieved from fits of the data using two different models (see main text for more details).}
    \label{tab:my_label}
\end{table}

In order to explain the results in SB$_{40}$, we now use a more complete analytical model (B), which includes both the 2s2p and 2p$^2$ resonances (see SM). The latter one is accessible via two-photon absorption due to the strong dipole coupling of the 2p$^2$ with the 2s2p state  \cite{JimenezPRL2014,JimenezGalan2016}. The 2p$^2$ state is also strongly coupled to the bound 1s2p state, which makes it necessary to include the non-resonant $1\text{s}^2\rightarrow1\text{s}2\text{p}\rightarrow2\text{p}^2$ transition \cite{JimenezGalan2016}. 
The new values of the $q$ parameter for the 2s2p resonance retrieved from the fit of model B to the experimental data are shown in Table \ref{tab:my_label}B. As expected, the results for SB$_{38}$ are the same as those obtained with model A since including the 2p$^2$ state, situated in the vicinity of SB$_{40}$, should not affect the lower sideband. In contrast, the real and imaginary parts of the $q$ parameter for SB$_{40}$ strongly differ from that of model A and are now very similar to those measured in SB$_{38}$. Figure \ref{Fig2}(e,f) shows the experimental data for SB$_{40}$, the fit (black dashed line) and the more complete model with $q=-2.77$ (red solid line), which are all in good quantitative agreement, indicating that the coupling between the 2p$^2$ state and the 2s2p resonance is responsible for the reduction in contrast and the smaller phase variation observed in SB$_{40}$ compared to SB$_{38}$. The calculations presented in the following use model B. 

\section{Discussion}

\subsection{Density matrix and entanglement}

Our aim is to understand how the 2p$^2$ state affects the coherence properties of the continuum wavepacket measured in SB$_{40}$. In the absence of any source of decoherence outside the atomic system, the wavepacket is a pure state corresponding to a coherent superposition of s and d waves [see Fig.~\ref{Fig1}(a)], which can be formally expressed as
\begin{equation}
    \ket{\Psi(t)}=\!\!\! \int\!\!\! d\epsilon\ c_s(\epsilon,t)\ket{R_s(\epsilon)}\otimes\ket{Y_{00}}+c_d(\epsilon,t)\ket{R_d(\epsilon)}\otimes\ket{Y_{20}},
    \label{Eq:4}
\end{equation}
where $\ket{R_{\ell}(\epsilon)}$ are the radial wavefunctions,  $\ket{Y_{\ell 0}}$ the spherical harmonics, $c_\ell(\epsilon)$ the coefficients of the coherent superposition, $\ell$ (= s or d) the electron angular momentum and $\otimes$ the tensor product. The integral is performed over the spectral bandwidth of the wavepacket, imposed by the excitation pulses. In general, the radial or angular wavefunctions are not separable, which implies that these two degrees of freedom are entangled \cite{HorodeckiRMP2009}. The quantum state of the wavepacket can equivalently be represented by its density matrix
\begin{equation}
\begin{split}
\nonumber
&\hat{\rho}(t)=\ket{\Psi(t)}\bra{\Psi(t)}\\
&= \!\iint\! d\epsilon d\epsilon' c_s(\epsilon,t)c_s^*(\epsilon',t)\ket{R_s(\epsilon)}\otimes\ket{Y_{00}}\bra{R_s(\epsilon')}\otimes\bra{Y_{00}}\\
&+c_d(\epsilon,t)c_d^*(\epsilon',t)\ket{R_d(\epsilon)}\otimes\ket{Y_{20}}\bra{R_d(\epsilon')}\otimes\bra{Y_{20}}\\
&+c_s(\epsilon,t)c_d^*(\epsilon',t)\ket{R_s(\epsilon)}\otimes\ket{Y_{00}}\bra{R_d(\epsilon')}\otimes\bra{Y_{20}}\\
&+c_d(\epsilon,t)c_s^*(\epsilon',t)\ket{R_d(\epsilon)}\otimes\ket{Y_{20}}\bra{R_s(\epsilon')}\otimes\bra{Y_{00}}.
\end{split}
\end{equation}

\begin{figure}
    \includegraphics[width=\columnwidth]{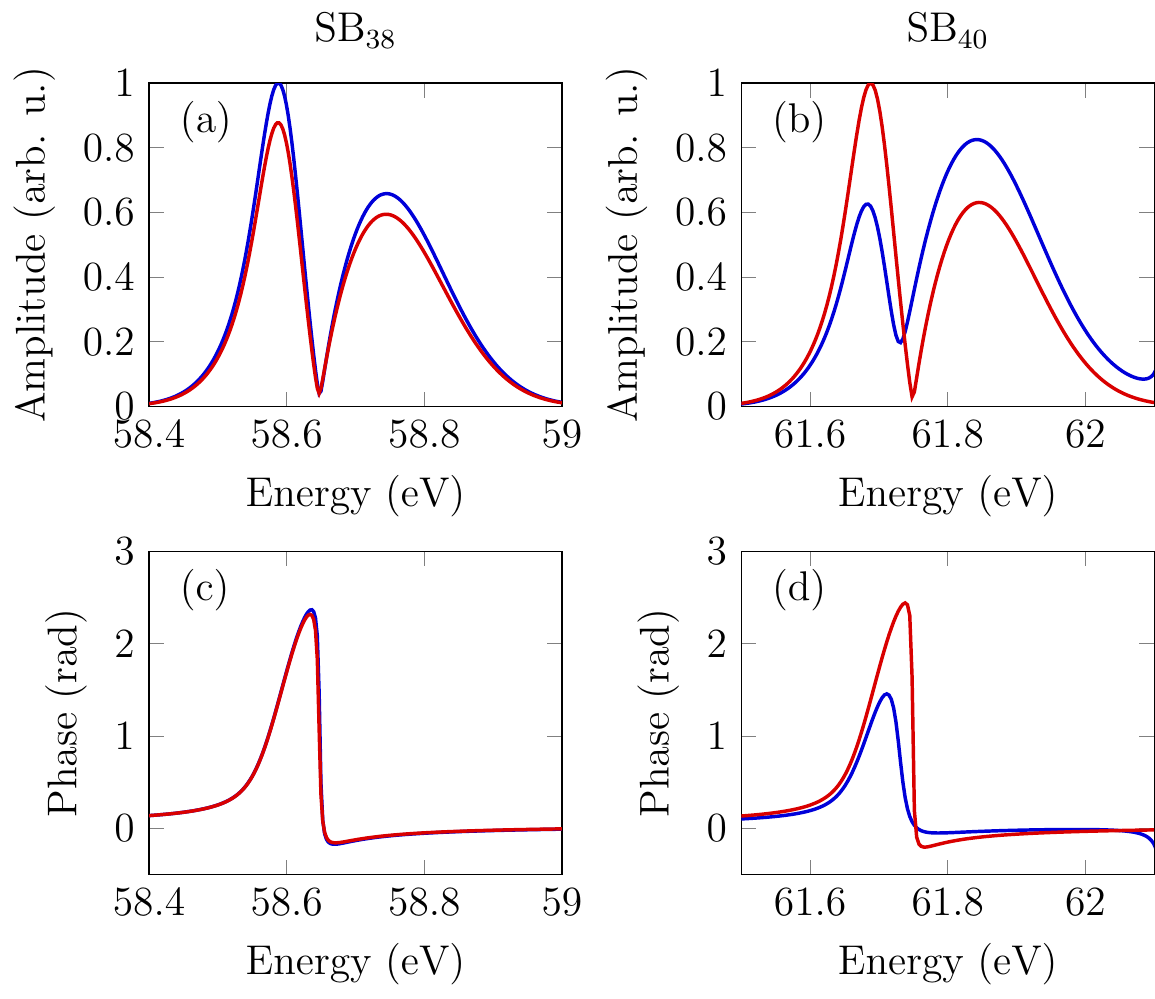}
    \caption{Resonant wavepackets in the s and d continua calculated using model B. Relative amplitude (top) and phase (bottom) of the s (blue) and d (red) wavepackets in SB$_{38}$ (a,c) and SB$_{40}$ (b,d).}
    \label{Fig4}
\end{figure}

The first two terms describe the reduced density matrices of the wavepackets in the s and d continua respectively, while the last two terms describe the coherences between the two continua. In the case of angle integrated measurements, as in this work, the angular degree of freedom of the photoelectron is not measured. As a result, the measured wavepacket is described by a reduced density matrix $\hat{\rho}_r$ defined as 
\begin{equation}\begin{split}
\hat{\rho}_r(t)&=\mathrm{Tr}_{(\theta,\phi)}[\hat{\rho}(t)]\\
 &=\!\iint\! d\epsilon d\epsilon'c_s(\epsilon,t)c_s^*(\epsilon',t)\ket{R_s(\epsilon)}\bra{R_s(\epsilon')}\\
&+c_d(\epsilon,t)c_d^*(\epsilon',t)\ket{R_d(\epsilon)}\bra{R_d(\epsilon')}.
\end{split}
\end{equation}
Due to the orthogonality of the spherical harmonics, the coherences between the different continua disappear and the final quantum state is a statistical mixture of the s and d wavepackets. In other words, when the radial and angular degrees of freedom are entangled, taking the partial trace over one of them leads to a mixed reduced density matrix characterized by lower degree of coherence compared to the pure case.

In the case of SB$_{38}$, the radial complex amplitudes for the s and d continua are almost identical, i.e. $c_s(E)\propto c_d(E) \propto A_\mathrm{R}(E)$ [see Fig.~\ref{Fig4}(a,c)]. As a consequence, it is possible to factorise the radial amplitudes such that the angular integration does not lead to a loss of coherence. In contrast, in SB$_{40}$, the 2p$^2$ state, which has $^1$S$_0$ symmetry, can only decay to the s continuum, leaving the d continuum unaffected (see Fig.~\ref{Fig1}). This results in the emission of s and d wavepackets with different amplitudes and phases as shown in Fig.~\ref{Fig4}(b,d). 
The amplitude of the d wavepacket is characterised by a strong destructive interference, with a maximum on the left almost twice as large as the right one. 
On the contrary, the amplitude of the s wavepacket shows a reduced interference contrast and the maximum on the right is slightly larger than the left one.
Similarly, the phase variation of the two wavepackets is different, with the d wavepacket showing a phase variation approximately twice as large as that of the s wavepacket. 
The radial wavefunctions for s and d waves are different so that they cannot be factorized in Eq.~(\ref{Eq:4}). As a result, the radial and angular degrees of freedom are entangled. Note that in Fig.~\ref{Fig4} we show the amplitude and phase of the resonant transition amplitude only (and not the phase difference between the non-resonant and resonant contributions as in the RABBIT scheme, see Fig.~\ref{Fig2}), so that the phases in Fig.~\ref{Fig4} (c,d) vary in the same way for the two sidebands.

   \begin{figure}
      \centering
      \includegraphics[width=\columnwidth]{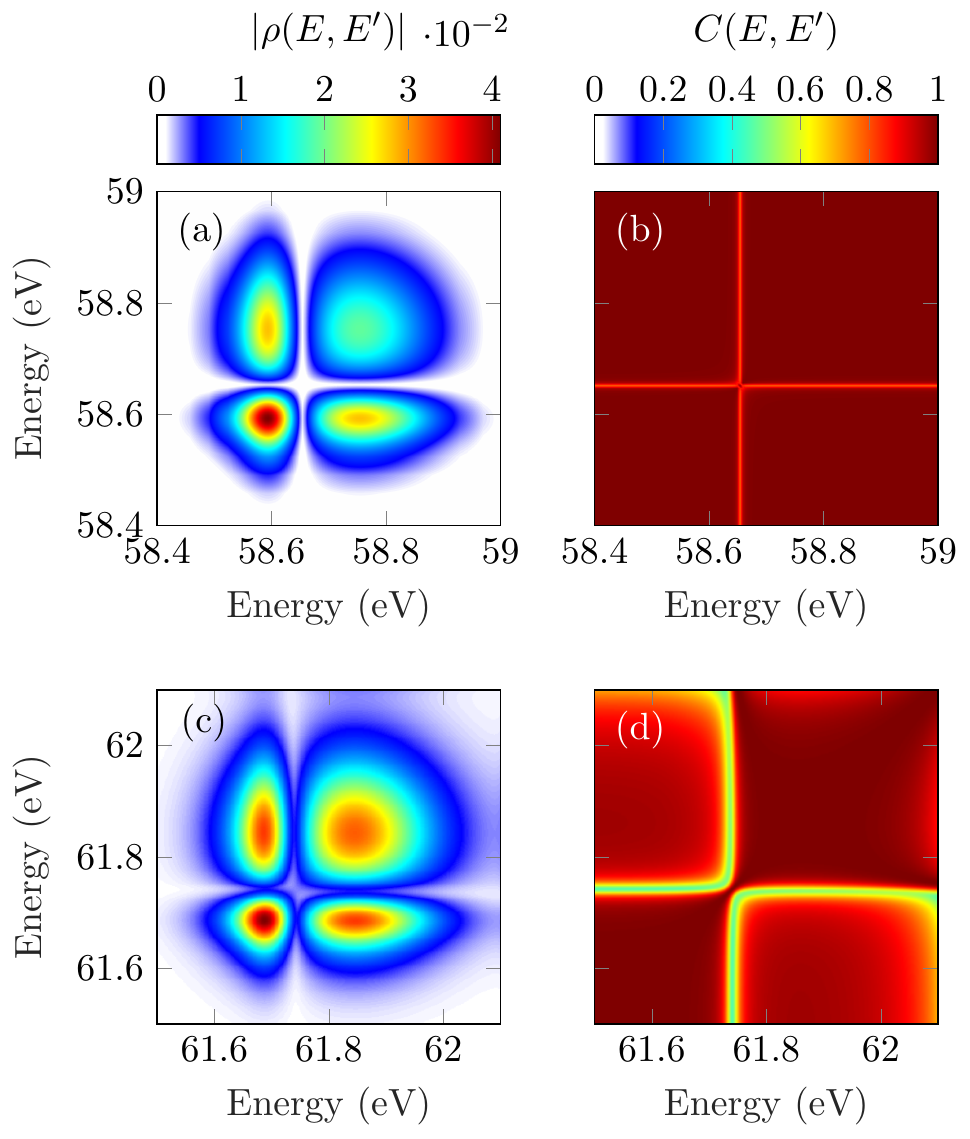}
      \caption{Quantum state of the asymptotic resonant two-photon wavepacket for SB$_{38}$ (a,b) and SB$_{40}$ (c,d). (a,c) Absolute value and (b,d) coherence map of the density matrix.}
      \label{Fig5}
  \end{figure}
  

\subsection{Decoherence}

We now investigate the asymptotic quantum state of the resonant two-photon wavepacket for the two sidebands, starting with SB$_{38}$. Figure \ref{Fig5}(a) shows the absolute value of the calculated reduced density matrix $|\rho_r(E,E')|=\lim_{t\to\infty}|\langle\phi_E|\hat{\rho}_r(t)|\phi_{E'}\rangle|$, where $\ket{\phi_{E}}$ are the asymptotic radial wavefunctions of any continuum state with well-defined  angular momentum and energy $E$. 
The populations, along the diagonal $\rho_r(E,E)$, present two maxima, separated by strong destructive interference at $58.65$~eV, reflecting the energy dependence of the photoelectron spectra [see Fig.~\ref{Fig2}(a)]. The off-diagonal elements, $\rho_r(E,E')$, which describe the coherences between the different final scattering states, are strongest when the corresponding populations $\rho_r(E,E)$ or $\rho_r(E',E')$ are high. To represent the degree of coherence between the scattering states at different energies, we introduce the coherence map $C(E,E')$ defined as
\begin{equation}
    C(E,E')=\frac{\left|\rho_r(E,E')\right|}{\sqrt{\rho_r(E,E)\rho_r(E',E')}},
\end{equation}
and shown in Fig.~\ref{Fig5}(b). Most of the wavepacket is fully coherent [$C(E,E')=1$], except for a slight decrease of the coherence in the vicinity of the destructive interference region. The purity of the wavepacket is extremely high with $\mathrm{Tr}(\hat{\rho}_r^2)=0.998$.

 \begin{figure}[t]
      \centering
      \includegraphics[width=\columnwidth]{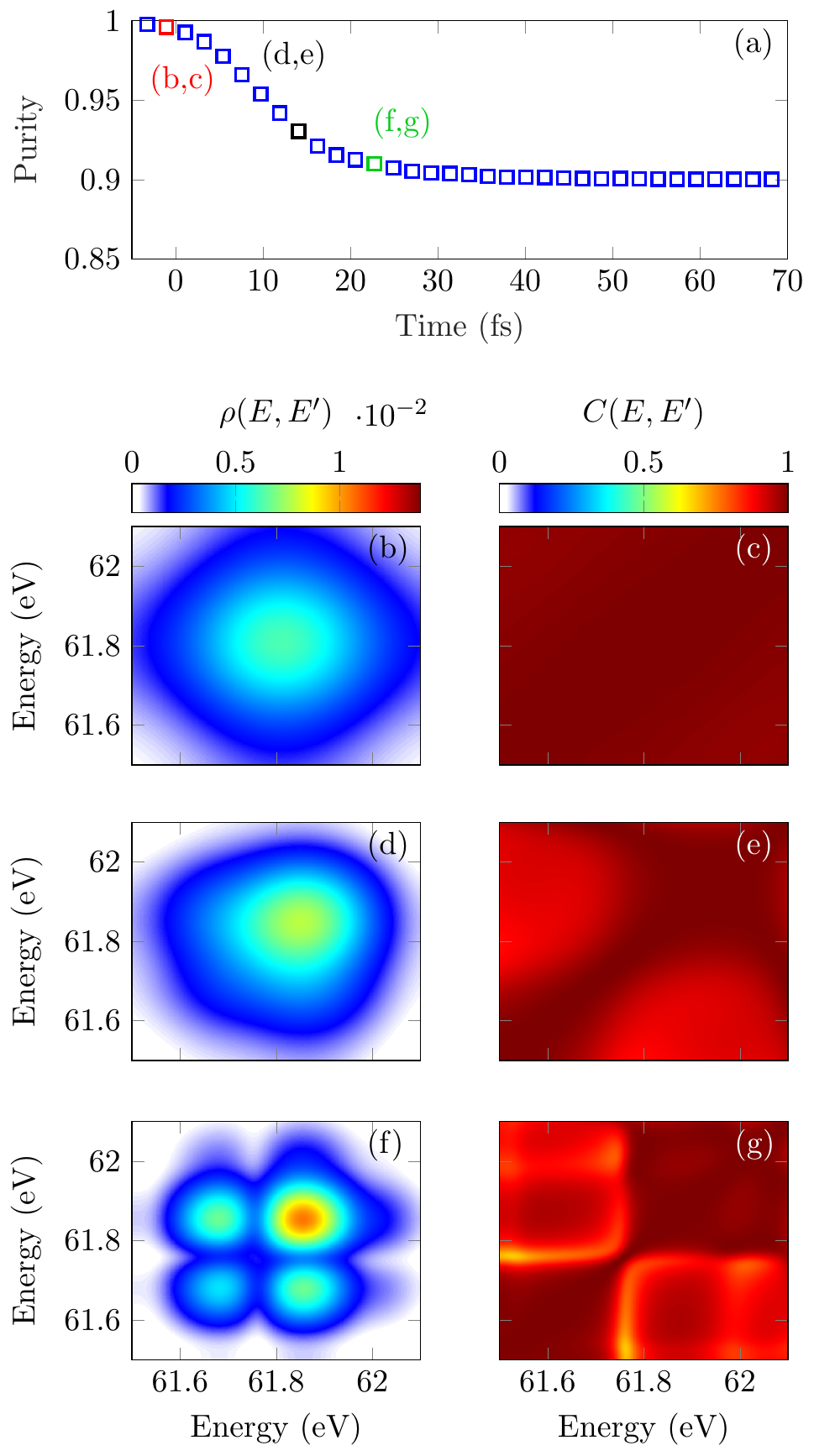}
      \caption{Quantum state evolution of the wavepacket in SB$_{40}$. (a) Temporal evolution of the wavepacket purity. (b-d) Absolute value of the density matrix at three different times as indicated in (a).  (e-g) Coherence maps corresponding to the density matrices in (b-d).}
      \label{Fig6}
  \end{figure}

Contrary to the case of SB$_{38}$, the resonant two-photon wavepacket in SB$_{40}$ is a statistical mixture of s and d waves with different amplitudes and phases as shown in Fig.~\ref{Fig4}(b,d). Knowing the theoretical amplitude and phase of the two angular components, we fit the relative weight of the two wavepackets in the measured complex amplitude, which allows us to reconstruct the density matrix. The s wavepacket is found to be 1.15 times stronger than the d wavepacket, which agrees well with the ratio of the angular coefficients for the p$\rightarrow$s and p$\rightarrow$d transitions, which is approximately 1.12. 
The absolute value of the asymptotic density matrix is shown in Fig.~\ref{Fig5}(c). Qualitatively, it looks similar to that obtained for SB$_{38}$, although the destructive interference is not as pronounced. 
In Fig.~\ref{Fig5}(d) we present the coherence map $C(E,E')$. 
The degree of coherence is lower than in the case of SB$_{38}$, between the low and high energy parts of the wavepacket and especially between the interference minimum and the rest of the wavepacket.
This originates from the difference between the s and d wavepackets in the region of the interference minimum, which are mixed in the angle-integrated measurement.  
Despite the clear loss of coherence between certain spectral regions of the wavepacket, the purity of the wavepacket remains high with $\mathrm{Tr}(\hat{\rho}_r^2)=0.89$. 

\subsection{Temporal evolution of the quantum state}

In the case of a pure wavepacket, the temporal profile and the buildup of the wavepacket as a function of time can be reconstructed using the simple transformation, \cite{GrusonScience2016,Busto2018,Turconi2020,DesrierPRA2018}
\begin{equation}
    A(E,t)=\int_{-\infty}^t\tilde{A}(t')e^{-\frac{iEt'}{\hbar}}dt',
\end{equation}
where $\tilde{A}(t)$ is the Fourier transform of the complex spectral amplitude $A(E)$. In the case of a mixed state, it is necessary to calculate the build up of the spectral amplitudes individually for the different angular channels in order to reconstruct the evolution of the density matrix and the coherence map during autoionization.

Figure \ref{Fig6} presents, for the SB$_{40}$ case, the evolution of the purity over time resulting from our model (a), as well as snapshots (b-g) of the density matrix and coherence map at three different times. At short times, the purity is close to 1 and both the density matrix and coherence map show that all the parts of the wavepacket are fully coherent. Indeed, around $t=0$, which corresponds to the maximum of the temporal amplitude of the wavepacket, direct ionization is the dominant ionization process so that the wavepackets in the s and d continua are almost identical, leading to a reduced density matrix corresponding to a pure state. As the 2s2p and 2p$^2$ states decay in the continuum, with lifetimes respectively  18\,fs and 112\,fs, the amplitude and phase of the s and d wavepackets become increasingly different, resulting in a loss of purity. This appears in the density matrix and coherence map as a loss of coherence between the low and high energy parts of the wavepacket. As time progresses, the destructive interference between direct ionization and autoionization appears in the density matrix and the degree of coherence between this spectral region and the rest of the wavepacket drops. The main decrease of purity occurs during the first 20~fs, which is close to the lifetime of the 2s2p resonance, while the coherences and populations keep evolving until approximately 50~fs, after which they have converged to the asymptotic value shown in Fig.~\ref{Fig5}(c,d) (see movie in SM).

In conclusion, we performed rainbow RABBIT measurements with high spectral resolution in He in the vicinity of the 2s2p Fano resonance. In the lower sideband, the emitted electron wavepacket is fully coherent while, in the upper sideband, the coupling between the 2s2p and 2p$^2$ resonances results in the emission of different wavepackets in the s and d continua, leading to an entanglement of the angular and radial degrees of freedom. Combining experiment and theory, we fully characterize the quantum state of the emitted wavepackets and show that this entanglement manifests itself as a loss of coherence in angle-integrated measurements. Finally, we reconstruct the quantum state evolution of the resonant two-photon wavepacket in SB$_{40}$ and monitor the degree of coherence of the wavepacket during autoionization. These results pave the way towards the complete characterization of complex partially coherent electronic wavepackets, extending the range of processes that can be investigated using attosecond photoelectron interferometry.

\section*{Acknowledgements}
The authors acknowledge Andreas Buchleitner and Christoph Dittel for fruitful discussions, and Alexandre Escoubas for support during the first experiments. DB thanks Eva Lindroth for providing non-resonant two-photon transition matrix elements and for discussions. The work was performed in the workframe of the European COST Action AttoChem. The authors acknowledge support from the Swedish Research Council (2013-8185, 2016-04907,2018-03731,2020-0520), the European Research Council (advanced grant QPAP, 884900) and the Knut and Alice Wallenberg Foundation. AL is partly supported  by the Wallenberg Center for Quantum Technology (WACQT) funded by the Knut and Alice Wallenberg foundation. DB acknowledges support from the Royal Physiographic Society of Lund and the Swedish Research Council (2020-2848). FM has been supported by the Spanish MICINN projects PID2019-105458RB-I00, the "Severo Ochoa" Programme for Centres of Excellence in R\&D (SEV-2016-0686) and the "María de Maeztu" Programme for Units of Excellence in R\&D (CEX2018-000805-M). LA acknowledges the NSF Theoretical AMO Grants No. 1607588 and No. 1912507. PS acknowledges Laserlab-Europe, Grant No. EU-H2020-871124, and Agence Nationale de la Recherche, Grants No. ANR-
20-CE30-0007-02-DECAP, No. ANR-11-EQPX0005-ATTOLAB, No. ANR-10-LABX-0039-PALM. DFS acknowledges funding from PAPIIT No. IA202821

\section*{Author contributions}
DB, CA, MI, SN, RJS, MT, SZ, CLA performed the experiments. DB and HL analyzed the experimental results with the help of DFS for the fit of the complex q parameters. DB, DFS, LA, FM, and AL elaborated the decoherence interpretation of the experiment and formalized it using the model developed by LA and FM. DB performed the numerical calculations. RJS and RF provided the high-resolution MBES and helped to optimize the spectral resolution. MG, PS, TP and AL supervised the project. All the authors discussed the results. DB wrote the article with the help of AL and with input from all the authors.

\section*{Methods}
\subsection{Experimental setup}
The Ti:Sapphire laser system delivers 30~fs, 3.5~mJ pulses at 1~kHz repetition rate. The central wavelength of the laser can be tuned between 790~nm and 810~nm. The pulses are sent into an actively stabilized Mach-Zender interferometer where 70\% of the pulse energy is directed in the pump arm of the interferometer where XUV attosecond pulse trains are generated. The IR pulses are focused using a spherical mirror with a 50 cm focal length into a neon gas cell, leading to the generation of high-order harmonics. The IR is then filtered out using an Al filter. The remaining 30\% of the initial pulse are sent into the probe arm of the interferometer, where the pulses are delayed with respect to the XUV attosecond pulse trains using a piezoelectric translation stage. A 10~nm bandpass filter centred at 800~nm is installed in the probe arm in order to reduce the spectral width of the probe pulse. The XUV and IR pulses are recombined using a drilled mirror that transmits the XUV and reflects the IR. The two pulses are then focused on an effusive helium gas jet using a toroidal mirror. The intensity of the IR is of the order of $10^{11}$~W/cm$^2$. The photoelectrons resulting from the interaction of the two pulses with the helium atoms are detected using a 2-m long MBES. Since the energy resolution of the spectrometer decreases with photoelectron kinetic energy, a retarding potential $U=32$~V is applied such that the lowest kinetic energy electrons that are detected are electrons from sideband 38.

\subsection{Deconvolution}
The use of narrowband probe pulses strongly reduces the influence of finite pulse (bandwidth) effects on the two-photon electron wavepacket. Hence, the main factor limiting the spectral resolution of the experiments is the spectrometer resolution. To circumvent this limitation, we use a blind iterative Lucy-Richardson deconvolution algorithm \cite{Richardson1972,Lucy1974}. Because the spectral resolution varies with electron kinetic energy, the different sidebands are deconvolved individually, assuming that the spectral resolution of the MBES is approximately constant across the sideband width (approx. 500 meV).

\subsection{Phase retrieval using Rainbow RABBIT}

The intensity of the sidebands neighboring the resonant photoelectron peak is described by Eq.~\ref{Eq:RABBIT}. The phase of the individual transition amplitudes can be decomposed as the sum of an XUV contribution  and an atomic response. In non resonant conditions the atomic phase is flat across the sideband, thereby providing a known reference, while the phase of the resonant two-photon transition is strongly influenced by the Fano resonance, with some smoothing due to finite pulse effects.

Both attosecond pulses and individual harmonics are intrisically chirped due to the generation mechanism. We distinguish between the effect of the attochirp, \textit{ i.e.} leading to a global phase variation from one harmonic to the other, and that of the femtochirp, affecting the properties of individual harmonics \cite{IsingerPTRSA2019}. The attochirp is constant for a given harmonic so that it only results in a global phase offset of the sideband phase. On the contrary, the femtochirp induces a phase variation across the harmonic width and may influence the measurements. For plateau harmonics, the chirp of consecutive harmonics is approximately constant so that, in the ideal case where the IR photon energy matches exactly half of the energy difference between consecutive harmonics, the femtochirp cancels and it does not affect the measurements. In practice,  the IR pulse used for generating the harmonics is blueshifted in the generation medium. As a result, the photon energy of the probe pulse does not match exactly half of the energy spacing between consecutive harmonics. In these conditions, the femtochirp from the consecutive harmonics does not cancel and leads to a linear phase variation across the sideband \cite{Busto2018,IsingerPTRSA2019}. We correct for this by fitting the linear phase in non resonant sidebands and subtract the linear slope in the measured sidebands.

\subsection{Amplitude retrieval using Rainbow RABBIT}

In order to extract the resonant amplitude, we take the Fourier transform of the oscillations and extract the amplitude at $2\omega$, ${\cal A}^{2\omega}_\mathrm{(N)R}(E)$ following the method described in \cite{GrusonScience2016,Busto2018}. For a resonant sideband, ${\cal A}^{2\omega}_\mathrm{R}(E)=2|A_\mathrm{R}(E)||A_\mathrm{NR}(E)|$, while for a non-resonant sideband, assuming that the non-resonant amplitude is similar for consecutive sidebands, ${\cal A}^{2\omega}_\mathrm{NR}(E)=2|A_\mathrm{NR}(E)|^2$. The resonant amplitude is hence given by
\begin{equation}
|A_\mathrm{R}(E)|=\frac{{\cal A}^{2\omega}_\mathrm{R}(E)}{\sqrt{2 {\cal A}_\mathrm{NR}^{2\omega}(E)}}.
\end{equation}
In practice, since ${\cal A}_\mathrm{R}^{2\omega}$ and ${\cal A}_\mathrm{NR}^{2\omega}$ are measured on different sidebands, ${\cal A}_\mathrm{NR}^{2\omega}$ must be shifted to the correct energy. Due to the blueshift, which cannot be measured accurately enough, it is difficult to determine how much ${\cal A}_\mathrm{NR}^{2\omega}$ should be shifted in energy. In order to determine the magnitude of the blueshift, we fit the $2\omega$ amplitude in sideband 38 with model A, where the blueshift is the free parameter. We can then apply the correct energy shift to the experimentally measured ${\cal A}_\mathrm{NR}^{2\omega}$. The same correction is applied to sidebands 38 and 40. The blueshift does not significantly affect the interference contrast in the amplitude but it affects the relative amplitude of the resonant and non resonant peaks in the sideband.

\subsection{Partially coherent fits}

In order to investigate the degree of coherence of the measured electron wavepackets, we simultaneously fit the measured amplitude and phase of the resonant wavepackets using a partially coherent finite pulse model. The finite pulse model depends on several parameters such as the harmonic width and the detuning from the resonance. Experimentally, it is difficult to determine these parameters with a precision better than $\sim $30~meV. 
The harmonic width is therefore fitted by comparing the experimental results for SB$_{38}$ with calculations using model A  (with $q$ real equal to -2.77). This value, which is found to be 142~meV is then fixed and kept the same for SB$_{38}$ and SB$_{40}$.
The determination of the energy detuning from the resonance is more difficult as it is very sensitive to the accuracy of the energy calibration. Since the link between time of flight and kinetic energy is not linear, a small error in calibration can result in slightly different apparent detunings for SB$_{38}$ and SB$_{40}$. In addition, the values returned by the fit for the real and imaginary parts of the $q$ parameter depend on the value of the detuning that is used. We circumvent this problem by performing the fit multiple times, randomizing the value of the detuning which can take values between 90~meV and 130~meV. The fits are repeated 50 times and each time the error returned by the fitting algorithm is saved. The values of $Re(q)$ and $Im(q)$ and the respective errors presented in the main text correspond to the values for which the fit residual is the smallest.

The model assumes Gaussian pulses. However, as mentioned in the main text, the MBES slightly distorts the lineshape, leading to a low energy  tail, which affects the quality of the fit. In addition, below a certain intensity threshold the measured phase is not reliable. In order to minimize this effects, we resample the measured amplitude and phase such that the center of the sideband is oversampled while the sides of the sideband are undersampled, thereby giving more weight to the Fano interference region. 

\subsection{Computing the Wigner distribution}

The Wigner distribution, and in particular the interference term, is very sensitive to phase noise. Below a certain threshold in the $2\omega$ amplitude, the phase extracted by the cosine fit is not reliable. However, in order to obtain high enough temporal resolution in the Wigner distribution it is necessary to select a spectral range that is large enough so that the $2\omega$ peak is fully covered. As a result, the phase originating from the wings of the $2\omega$ signal can strongly affect the Wigner distribution. In order to suppress this effect, in Fig.~\ref{Fig3}, we set the phase that is measured below a threshold of 35$\%$ of the maximum 2$\omega$ signal to a constant value. The noise in the $2\omega$ amplitude does not significantly distort the Wigner distribution such that no filtering is needed in amplitude.
\bibliography{Ref_lib}

\end{document}